\documentstyle[12pt]{article}
\def \beq{\begin{equation}}
\def \eeq{\end{equation}}
\def \pa{\partial}
\newcommand{\n}{
\newpage
\pagestyle{empty}
\noindent
}
\begin{document}
\n
\begin{center}
\large

{\Large\bf  Macroscopic   Einstein   equations   for   a   system   of
interacting particles and their cosmological applications.}

\bigskip
     A. V. Zakharov

\medskip
{\it Dept.Gen.Rel.\& Grav.\\
     Kremljevskaja Str. 18\\
     Kazan Stata University\\
     420008 Kazan, Russia\\
     tel.:(+7)(8432)315586\\
     e-mail: Alexei.Zakharov@ksu.ru}

\bigskip

     {\bf SUMMARY}

\end{center}

One of the possible applications of macroscopic  Einstein equations
has been considered. So, the nonsingular isotropic and uniform
cosmological model is built. The cosmological  consequences  of this model
are agree with conclusions  of standard  hot model of the Universe.

\newpage

\pagestyle{plain}

\setcounter{page}{1}

\begin{center}
{\Large \bf  Macroscopic   Einstein   equations   for   a   system   of
interacting particles and their cosmological applications.}

\bigskip

     A. V. Zakharov
\end{center}

\bigskip

{\bf 1.  Macroscopic system of Einstein equations}

\bigskip

As it is known the macroscopic Maxwell equation for continuous media  can
be obtained  also from  the  microscopic  Maxwell  equations  by  ensemble
averaging the latter (refer to.[1]-[3]).

The Einstein equations,  which right-hand sides contain the
energy-momentum tensor  of  matter,  are  phenomenological  equations.
It is
natural to  suppose   that   the   Einstein   equations   (or   their
generalizations) for  continuous  media  can  be also obtained from the
microscopic Einstein equations, i.e., Einstein equations which
right-hand side  contain  the  sum  of  the  energy  -  momentum tensors of
individual particles.  However,  due to the nonlinearity of the
Einstein equations left-hand side,  the averaging of the
microscopic Einstein
equations is more complicated than one of the  microscopic  Maxwell
equations (refer to. [4] -  [6]).

     In Ref. [7] - [9] were obtained  the  macroscopic  Einstein equations
for a  system of interaction particles up to second order in the
interaction constant. They have the forms:
     \beq \label {1}
     G_{ij} + Z_{ij} = \chi T_{ij},
     \eeq
     Here $G_{ij}$ is the Einstein's tensor of  the  Riamannian  space
with  macroscopic  metric  $g_{ij}$, $T_{ij}$ is the macroscopic
energy - momentum tensor,  $\chi = 8\pi k/c^4$ is  Einstein  constant,
$k$ is the gravitational constant, $c$ is the velosity of light.

The Einstein equations of the gravitational field for continium media
differ from the classical Einstein equations  by
the presence of additional tensor $Z_{ij}$
in the left-hand side. It caused by particle interaction.

The form  of  this  tensor  is  given  in  Ref.[8]  in  the  case   of
gravitationally interaction  of particles and in Ref.[9] in the case
of electromagnetically interaction of particles.

     The tensor $Z_{ij}$ is the traceless tensors with zero divergence:
     \beq \label {2}
     g^{ij} Z_{ij} = 0, \qquad Z^{ij}_{;j}=0.
     \eeq
where the semicolon denote a covariant derivative in a spase  with
the macroscopic metric $g_{ij}$.

     In the  case  of gravitationally interacting particles the tensor
$Z_{ij}$ has the form (Ref. [8]):
     \beq \label {3}
    Z_{ij} =   \varphi^k_{ij;k}    +   \mu_{ij},
     \eeq

     where
$$\varphi^{k}_{ij} = - \sum_{bc} \frac {\chi^{3} m_{b}^{3} m_{c}^{3}
c^9 }{8 (2 \pi)^3}\int \frac {d^3 p'}{p'^0 \sqrt {(-g)}}\int
\frac {d^3 p''}{p''^0 \sqrt {(-g)}}{\biggl[}
\frac{1}{2} g^{fk} u''_{i} u''_{j} + $$
$$+u'^{k}  (u'  u'')(\delta^{f}_{j}  u''_{i}  +  \delta^{f}_{i}
u''_{j}){\biggr]}\left( (u' u'')^2 - \frac {1}{2}\right) $$
     \beq \label {4}
 K_{f\alpha}(u', u'')\left( F''_c \frac {\pa F'_b}{\pa p'_{\alpha}} -
F'_b \frac {\pa F''_c}{\pa p''_{\alpha}}\right),
     \eeq
$$ \mu_{ij} = - \sum_{bc} \frac {\chi^3 m_{b}^{3} m_{c}^{3} c^9}
{8 (2 \pi)^3}\int \frac {d^3 p'}{p'^0 \sqrt {(-g)}}\int
\frac {d^3   p''}{p''^0 \sqrt {(-g)}}{\biggl\{} {\Bigl[} (z^2 + \frac {1}{2})
(u''_i u''_j + u'_i u'_j) + $$
$$ + (z^2 - \frac {1}{2}) g_{ij} - 2z (u'_i u''_j + u''_i u'_j){\Bigr]}
g^{qr}- 2(z^2- \frac {1}{2}) \delta^{q}_{i} \delta^{r}_{j}{\biggr\}}$$
     \beq \label {5}
F''_c \frac {\pa}{\pa p'_f}{\biggl\{} F'_b{\Bigl[} (z^2-
\frac {1}{2})\delta^{m}_{f} + (z^2 + \frac {1}{2}) u'_{f} u'^m -
2z u''_{f} u'^m {\Bigr]}{\biggr\}}J_{rqm}(u', u'') .
     \eeq

     Here $p'^i$ is the momentum of particles of species $b$,  $p''^i$
is the  momentum  of  particles  of  spesies  $c$  $u'^i=p'^i/m_b  c$,
$u''^i=p''^i/m_c c$,  $F'_b$  is  the  one  - particle distribution
function of particles of spesies $b$, $F''_c$ is the one  -  particle
distribution function of spesies $c$,  $z=u'^i u''_i$, $m_b$ and $m_c$
are the mass of particles of spesies $b$ and $c$ respectively,
$$\frac  {d^3  p'}{p'^0  \sqrt  {(-g)}} \hspace{1cm}
and \hspace{1cm}\frac  {d^3  p''}{p''^0  \sqrt  {(-g)}}$$
are the invariant volume elements in tree - dimensional momentum space
of particles spesies "b" and "c" respectively.

The greek index $\alpha$ in (\ref {4}) takes
the values 1,2 and 3 only (the spartial index).  The derivative with respect
to $p'_f$ in (\ref {5}) should be calculated as all four  components
of momentum are independent. The dependence of $p'_0$ on $p'_{\alpha}$
is taken into account after differentiation with  respect  $p'_f$
is completed.

The tensors  $K_{ij}(u',  u'')$  and $J_{ijk}(u',  u'')$ have the form:

$$ K_{ij}(u', u'')    =    \frac    {4    \pi^2}{k_{min}^{2}[(u'u'')^2 -
1]^{3/2}}{\bigl\{} - {\lbrack} (u' u'')^2 -1 {\rbrack} g_{ij} - $$
     \beq \label {6}
- u'_{i} u'_{j} - u''_{i} u''_{j} + (u' u'')(u'_{i} u''_{j} +
u''_{i}  u'_{j}){\bigr\}}
     \eeq
 $$ J_{ijk}(u',u'') = A {\Bigl[} (g_{ij}u'_k +g_{ik}u'_j + g_{jk}u'_i) -
z(g_{ij} u''_k +g_{ik} u''_j + g_{jk} u''_i) - $$
$$ - (u'_i u''_j u''_k + u''_i u'_j u''_k + u''_i u''_j u'_k) +
3z u''_i u''_j u''_k {\Bigr]} + $$
$$ + C {\Bigl[} u'_i u'_j u'_k - z (u'_i u'_j u''_k + u'_i u''_j u'_k +
u''_i u'_j u'_k) + $$
     \beq \label {7}
 + z^2 (u'_i u''_j u''_k + u''_i u'_j u''_k + u''_i u''_j u'_k) -
 z^3 u''_i u''_j u''_k {\Bigr]},
     \eeq
     where
     \beq \label {8}
 A  =  -  \frac  {2\pi  \sqrt  {2}}{k_{min}}{\biggl[} \frac {(z-2)}{(z-1)^2
(z+1)^{1/2}} + \frac {(2z-1)}{(z+1)(z-1)^{5/2}} \ln \left( z +
\sqrt {z^2 - 1}\right) {\biggr]},
     \eeq
     \beq \label {9}
 C = - \frac {2\pi \sqrt {2}}{k_{min}}{\biggl[} \frac {(z-6)}{(z-1)^3
(z+1)^{3/2}} + \frac {(6z-1)}{(z+1)^2 (z-1)^{7/2}} \ln \left( z +
\sqrt {z^2 - 1}\right) {\biggr]}.
     \eeq

In these expressions
     $$k_{min} = \frac {1}{r_{max}},$$
     where $r_{max}$ is the size of the correlation region in the case
of gravitationally interacting particles. In Refs.
[10], [11] there are estimates for $r_{max}$ in the case where
the average metric  $g_{ij}$
is the metric of isotropic cosmological model.

 \bigskip

{\bf 2. Applications of the theory. Nonsingular isotropic and uniform
 cosmological models in macroscopic theory of gravity.}

\bigskip

Let's consider received equations for the ambience, base at
local thermodynamic  balance.  In this cause the distribution function
of each sorts of particles has the type ((refer to. [12]):

  \beq \label {40}
     F_a( p_{\alpha}) = A_a exp [-c(v_i p^i)/(k_B T)].
     \eeq

Here $v_i$ is the macroscopic four-velocities of ambience, $k_B$ is the
Boltzman's constant, $T$ is the temperature, $A_a$ is the normalizing
constant.

In this case tensor $  \varphi^{k}_{ij}  $ becomes zero,
but tensor $\mu_{ij}$  has  the form:

  \beq \label {55}
     \mu_{ij} =  \chi  \epsilon_1 \left(\frac   {4}{3}v_i   v_j   -   \frac
{1}{3}g_{ij}\right)
     \eeq

This is correct both for gravitational and electromagnetic
interactions.

If move $\mu_{ij}$  from left-hand side to the right-hand side ,
the macroscopic equation
effectively change into usual Einstein equations with the additional
tensor of energy-momentum  in the right-hand side. Moreover the "additional
tensor of energy-momentum"  has  the form of energy-momentum tensor
of ideal liquid with the state equation  $P_1 = \epsilon_1/3$,
but with negative "density
of energy" $(-  \epsilon_1)  $ and negative "pressure" $(-P_1)$ .
Via $\epsilon_1$  and  $P_1$ their absolute values are marked.

In the case of gravitational interactions in nonrelativistic limit
when $mc^2  >> k_B T $ , the absolute value of this "negative density
of energy"
is
  \beq \label {56}
 \epsilon_1 = \sum_{ab} \frac{4 k^2 r_{max}}{ k_B T}
m_{a}^{2} m_{b}^{2} N_a N_b ,
     \eeq
where $N_a$ and $N_b$ is the number of particles of  species
$a$ and $b$ recpectively densities, $m_a$ and $m_b$ are their masses.

It is proportional to the square of gravitational constant and square of
number of particles density.
Consequently the additional terms can play significant role in ambiences
of sufficiently high density.
Such density is possible on the early stages of the Universe evolution. So,
naturally, one can use the first exhibits of the received equations in
the theories of the early stages of the Universe evolution.

Thereby there is a real possibility of manifestation an interaction on
early stages of the Universe evolution.

Let's turn to the building of uniform and isotropic cosmological
models within the framework of macroscopic theory of gravity.

Let's write the metricses of these models in a form:
  \beq \label {57}
     (d s)^2 = a^{2}(\eta) \left( (d\eta)^2 - (dr)^2 - \phi^{2}(r)
((d\theta)^2 + \sin^{2} (\theta)(d \varphi)^2)\right).
     \eeq
Here $\phi (r) = r$ , $\phi (r) = \sin r $, $\phi (r) = \sinh r $
for the flat, close and open models respectively.

The system of Einstein equations for these metricses is reduced to two
equations (refer to ,for instance, [13]):

  \beq \label {58}
    a'^{2} + \xi a^2 = \frac {8\pi k}{3 c^4} a^4 \tilde \epsilon,
\eeq

  \beq \label {59}
\frac {d \tilde \epsilon}{\tilde \epsilon + \tilde P} = - 3\frac {d a}{a}.
  \eeq

   Here stroke under $a$ marks derivative on time variable
$\eta$; $\tilde \epsilon $  and  $\tilde  P$ is a density of energy
and pressure of matter , $\xi = 0,+1,-1$ for the flat,
closed and open models respectively.

The macroscopic Einstein equations are distinguish from (\ref {58}) -
(\ref {59}) by following.  In the first, $\tilde \epsilon $ in the
right-hand side of (\ref  {58}) is replased by the
difference of usual energy density $\epsilon$ and absolute value
"density of energy" $\epsilon_1$, stipulated by the particles interaction
of ambience.

    At the present moment of the Universe evolution
the expression (\ref {56}) for $\epsilon_1$ is correct.

In the second, the equation (\ref {59}) is changed by two similar:

  \beq \label {60}
\frac {d \epsilon}{ \epsilon +  P} = - 3\frac {d a}{a},
   \eeq

  \beq \label {61}
\frac {d \epsilon_1}{  \epsilon_1 + \ P_1} = - 3\frac {d a}{a}.
   \eeq

  Here $P$ is a usual pressure of matter,
$ P_1  =  \epsilon_1/3 $ is the "pressure", stipulated by
  the interaction of particles.

Equations (\ref  {60}) and (\ref  {61}) follows from the fact, that
divergency of the usual  energy - momentum tensor and divergency of the
"additional energy - momentum tensor",  stipulated by the interaction,
equals to zero.

From (\ref {56}) it follow that in present the value $\epsilon_1$ and
the relict radiation energy density are of the same order of value
(see to below).
So, if we take into account a contribution $ \epsilon_1 $ to
$\tilde \epsilon $, we must
take into account a contribution of  the  relict  radiation  energy
density to $\tilde \epsilon $ density.

So, let's put in the right part of (\ref {58})

  \beq \label {62}
     \tilde \epsilon = \epsilon_{m} - \bar \epsilon.
   \eeq

 Here $\epsilon_m$ is the density of energy of material disregarding
density of energy
of relict radiation, but under $\bar \epsilon  $ we shall understand
a difference
between the absolute value "density of energy", stipulated by the
interaction, and density of energy of relict radiation. This is
suitable, since equations of condition for relict radiation and for
"energy - momentum tensor", stipulated by the interaction alike:
pressure is one  third from density of energy.
So that the equation (\ref {60}) and (\ref {61}) coused from the
following $\bar \epsilon $ dependency
 on the scale factor:

  \beq \label {63}
\bar \epsilon  = \frac {3 c^4 a_{1}^{2}}{8\pi k a^4}.
  \eeq

Here $a_{1} = const $.

For density of energy of rest material from (\ref {60}) we have:

  \beq \label {64}
 \epsilon_m  = \frac {3 c^4 a_{0}}{4\pi k a^3},
  \eeq

where      $ a_{0} = const $ .

We substitute  (\ref {62}), (\ref {63})
and (\ref {64}) into (\ref {58}).  As a result we
obtain the equation:

  \beq \label {65}
    a'^{2} + \xi a^2 = 2a_{0}a - a_{1}^{2}
     \eeq

The solution  of  this equation for the scale factor $a(\eta)$ one can
write in the form:

  \beq \label {66}
   a = a_{m} + \frac{1}{2} a_{0}\eta^2
  \eeq
for flat ($\xi = 0$) models ( here $a_{m}= a_{1}^{2}/2a_{0}$),

  \beq \label {67}
   a = a_{m} + (a_{0} - a_{m})(1-\cos \eta)
  \eeq
     for closed  ($  \xi  =  +1  $)  models  (here $a_{m} = a_{0} -
\sqrt{(a_{0}^{2} - a_{1}^{2})} $), and

  \beq \label {68}
   a = a_{m} + (a_{0} + a_{m})(\cosh \eta - 1)
   \eeq
   for open ($  \xi  =  -1  $)  models (here $a_{m} =
\sqrt{(a_{0}^{2} + a_{1}^{2})} - a_{0}$ ).

As it is seen, all three models are nonsingular: under $\eta = 0$
the scale factor $a(\eta)$
does not apply to the zero as in standard cosmologycal models, but
takes a minimum value $a_{m}$.

Let us calculate a value of density  $  \rho_{m}$  of  nonrelativistic
matter at the
moment, when scale factor takes a minimum value $a_m$:

  \beq \label {69}
      \rho_{m} = \rho_0 \frac {a^3(\eta_0)}{a^{3}_{m}}.
\eeq

Here $\rho_0  $ is the value of usual matter density at present moment.

This moment corresponds a time coordinate value $\eta_0$.
Substituting
(\ref {66}) --- (\ref {68}) into (\ref {69}) we  obtain the expression
for $\rho_{m}$ via $\rho_{0}$, $\eta_0$ and ratio $a_0/a_m$.

Let us write down the expressions for
$\Omega = (\epsilon_m - \bar \epsilon)_{\eta=\eta_0}/(c^2\rho_{кр})$,
and $\lambda  =  (\bar \epsilon/ \epsilon_m)_{\eta=\eta_0}$
via $\eta_0$ and ratio $a_0/a_m$.

Here $\rho_{кр} = 3 H^{2}_{0}/8\pi k $ is the critical density at
the present moment, $\Omega$ is the densities parameter, $\lambda$ is a
ratio of  "energy  density",  stipulated  by  the interaction absolute
value, and density of relict radiation energy difference to the density
of usual matter energy at the present moment.

From (\ref {58}), (\ref {62}) - (\ref {64}) we obtain
  \beq \label {70}
\Omega = \left(1+\xi \frac{a^2}{a'^2}\right)_{\eta=\eta_0} \qquad
\lambda = \frac {a_{1}^{2}}{2a_0 a(\eta_0)}
\eeq

Substituting
(\ref {66}) --- (\ref {68}) into (\ref {70}) we  obtain the expression
for $\Omega$ and $\lambda$ via, $\eta_0$ and ratio $a_0/a_m$.

Expressing from received correlations $\eta_0$ and  $a_0/a_m$ via
$\Omega$  and $\lambda$  and then substituting then received expressions
for $\eta_0$ and  $a_0/a_m$ in (\ref  {69}) ,
we obtain the following result:

  \beq \label {71}
   \rho_m = \frac{\rho_0}{\lambda^3}\left(\frac{1}{2} +  \sqrt  {\frac
{1}{4} - \lambda (1-\lambda)\frac {(\Omega -1)}{\Omega}}\right)^{3}
    \eeq

   Result (\ref {71}) is correct for flat $(\Omega = 1)$ , open
$(\Omega < 1)$, and closed $(\Omega > 1)$ models.

The received models can be realized, if absolute value of "energy density"
(\ref {56}), stipulated by the interaction, exceeds at
present moment the energy density of relict radiation.

The  "energy density"
(\ref {56}) , stipulated by interaction,
is created basically galaxies accumulations with the mass
$m \sim  10^{15} - 10^{16}M_{\odot}$. This  "energy density"
is estimated as
$$ \frac {12 k^2 t_0 \rho^2_0 m}{<v>}. $$

When getting this correlation we put $3 k_B T =  m<v>^2$,
where $<v>$ is an average chaotic velocity of galaxies accumulations,
$N = \rho_0/m$ , but parameter $r_{max}$ is estimated as
$<v>t_0$, where $t_0$ is the modern age of the Universe
(refer to. [10], [11]).

Substituting into  the obtained ratio the numerical values of fundamental
constants, cosmological time, average density of matter in the Universe
at present moment (refer to ,for instance, [14]), we obtain under
$<v>/c = 10^{-6}$, $m \sim  10^{15}M_{\odot}$:

$$ \epsilon_1 \sim 7 \cdot 10^{-13} erg/cm^3$$

It is approximately the value of relict radiation energy density
we have at present moment
(refer to.[14]).

Consequently, realization of such situation is possible, when absolute
value of  "energy  density",  stipulated  by  interaction,  exeeds the
density of relict radiation energy.
And it is the case considered in the paper when $\lambda > 0$.

On initial stage of the Universe evolution the density
$\rho_m    \sim \rho_0/\lambda^3 $  is
sufficiently great , so usual scenario of the Universe hot models preserves.
For a  moment  when scale factor takes a minimum value and the density
of matter is maximum let us introduse the folloing.

If at this moment a temperature does not exceed the temperatures of
nucleon - antinucleons couple annihilation, the density of
nonrelativistic mater can be evaluated as

     $$ \frac {m_p 10^{-8}\sigma T^4}{k_B T}, $$
where $ \sigma$  is the constant of Stefan - Boltzman, $m_p$ is
the mass of proton.

Comparing this expression with $\rho_0/\lambda^3 $ one can make a conclusion,
that minimum
value of scale factor will be reach on leptones stage of the Universe
evolution when parameter $\lambda$ is within
$10^{-10}$ - $10^{-12}$, and the temperature will
be about $10^{10}K^o$ - $10^{12}K^o$.

 \bigskip

{\bf 3. Domain of the theory application.}

\bigskip

     When concluding the macroscopic Einstein equations we suggest the
distant collision to be the prevalent. It is correct if
     $$\gamma = \frac {E_p}{E_k} \ll 1,$$
     where $E_p$ is the potential energy of two particles interaction,
     $E_k$ is the kinetic energy of particles.

     For the  system of electromagnetical interacting particles in a
condition of local thermodynamic balance we can put $E_k \sim  kT$,
$E_p \sim e^2 n^{1/3}$.

     So the condition of validity takes the form
  \beq \label {72}
     \gamma = \frac {e^2 n^{1/3}}{kT} \ll 1.
     \eeq

     Here $e,  n,  T,  k$ is  the  charge,  density,  temperature  and
Boltzman constant correspondingly.

     Let us  verify  (\ref  {72}) in constructing model under $t=t_m$,
when the  Universe  reaches  the  maximum  density.  The  majority  of
electromagnetical interacting  particles  on  this stage compose the
electron-positron pairs.  It's density is compared  on  the  order  of
value with  the  one of relict photons and can be estimated as $\sigma
T^4/kT =  \sigma  T^3/k$,  where  $\sigma$  is  the   Stephan-Boltzman
constant. Then for parameter $\gamma$ we have estimation:
     $$ \gamma = \frac {e^2 \sigma^{1/3}}{k^{4/3}}  \sim
10^{-2}  \ll  1.$$
     Consequently the  application  condition   of   the   macroscopic
Einstein equations in this model is not violated.


\begin{thebibliography}{14}
\bibitem{1}
H. A.  Lorentz.  Versush  einer  Theorie  der Elektrishen und Optishen
Ercheinungen in Bewegten Korpern. Leiden, (1895)

\bibitem{2}
H. A. Lorentz. The Theory of Electrons. Leipzig: Teubner, (1916)

\bibitem{3}
Yu. L.  Klimontovich,  Kinetic Theory of Electromagnetic Processes [in
Russian],Nauka, Moscow (1980)

\bibitem{4}
Shirokov M.  F.,  I.  Z.  Fisher, Astron. Zh. {\bf 39}, 899 (1962) [in
Russian]


\bibitem{5} G.  F. R. Ellis. General Relatyvity and Gravitation, ed B.
Bertotti, F. de Felici and a> Pascolini. Dortrecht: Reidel. (1984)

\bibitem{6}
R. M. Zalaletdinov, Gen Relativ. Gravit. {\bf 25},673 (1993)

\bibitem{7}
A. V.  Zakharov,  Zh.  Eksp. Teor. Fiz. {\bf 110}, 3 - 10 (1996) [Sov.
Phys. JETP {\bf 83}, 1 - 8, (1996)].

\bibitem{8}
A. V.   Zakharov,   Zh.  Eksp.  Teor.  Fiz.  {bf  112},  1153  -  1166
(1997) [Sov. Phys. JETP {\bf 85}, 627 - 634 (1997)].

\bibitem{9}
A. V.  Zakharov.  Macroscopic  Einstein  equations  taking into account the
electromagnetic interaction of particles of  medium.//In  :  Noveishie
problemi teprii polya. Edit: prof. Aminova A.V., Kazan 1998 p.118-124
(in Russian)

\bibitem{10}
G. S. Bisnovatij - Kogan, I. G. Shukhman, Zh. Eksp. Teor. Fiz.
{\bf 82}, 3 (1982).

\bibitem{11}
A. V. Zakharov, Zh. Eksp. Teor. Fiz. {\bf 99}, 769 - 782 (1989). [Sov. Phys.
JETP, {\bf 69}(3), 437 - 443, (1989)].

\bibitem{12}
N. A. Chernikov, Nauchn. docl. visch. shcol. Fiz. mat. N 1, 168 (1959).

\bibitem{13}
L. D. Landau, E. M. Lifshitz. Teorija polja. Izd-vo "Nauka" M., (1973).
(in Russian)

\bibitem{14}
Ja. B. Zeldovich, I. D. Novikov. Stroenie i evoljutzija Vselennoi.\\
Izd-vo "Nauka" M., (1975).(in Russian)

\end{thebibliography}
\end{document}